\documentclass{aa}  
\usepackage{graphicx}
\usepackage{txfonts}
\usepackage{natbib}
\begin{document}
   \title{A synthetic stellar polarization atlas from 400 to 1000 nm}

%   \subtitle{I. Overviewing the $\kappa$-mechanism}

   \author{H. Socas-Navarro\inst{1},
          A. Asensio Ramos\inst{2}
          \and
          R. Manso Sainz\inst{2}
          }

   \offprints{H. Socas-Navarro}

   \institute{High Altitude Observatory, NCAR\thanks{The National
       Center for Atmospheric Research (NCAR) is sponsored by the
       National Science Foundation.}, 3080 Center Green Dr, Boulder CO
       80301, USA\\
       \email{navarro@ucar.edu}
         \and
	 Instituto de Astrof\' \i sica de Canarias, Avda V\' \i a L\'
       actea S/N, La Laguna 38200, Tenerife, SPAIN
             }

         \date{}
%   \date{Received September 15, 1996; accepted March 16, 1997}

\newcommand{\citeN}[1]{\citeauthor{#1} (\citeyear{#1})}
\newcommand{\citeNP}[1]{\citeauthor{#1} \citeyear{#1}}
\newcommand{\citeyearNP}[1]{\citeyear{#1}}
\newcommand{\citeANP}[1]{\citeauthor{#1}}

%% --- Some useful spectral definitions

\newcommand{\CaII}{\ion{Ca}{2}}
\newcommand{\NaI}{\ion{Na}{1}}
\newcommand{\FeI}{\ion{Fe}{1}}
\newcommand{\HeI}{\ion{He}{1}}

% \abstract{}{}{}{}{} 
% 5 {} token are mandatory
 
  \abstract
   {With the development of new polarimeters for large telescopes, the
  spectro-polarimetric study of astrophysical bodies is becoming
  feasible and, indeed, more frequent. In particular, this is
  permitting the observational study of stellar magnetic fields}
   {With the aim to optimize and interpret this kind of observations, 
  we have produced a spectral atlas of circular polarization in a grid
  of stellar atmospheric models with effective temperatures between
  3500 and 10000~K, surface gravities $\log(g)=3.5-5$, metallicities
  between 10$^{-2}$ and 1, and magnetic field strengths of 100, 1000
  and 5000~G}
   {We have computed the emergent Stokes~$I$ and~$V$ flux spectra in LTE
  of more than 10$^5$ spectral lines}
   {The atlas and several numerical tools are available in electronic format and may be
  downloaded from http://download.hao.ucar.edu/pub/PSA/. 
%There are also
%  IDL, Fortran~77, Fortran~90 and C procedures that illustrate how to
%  read and use the files. 
  In this paper we review and discuss some of its most relevant
  features, such as which spectral regions and individual lines harbor
  the strongest signals, what are interesting lines to observe, how to
  disentangle field strength from filling factor, etc.}
   {}

   \keywords{atlases --
             Stars: atmospheres --
             Stars: magnetic fields --
             Sun: magnetic fields
            }

   \maketitle
%
%________________________________________________________________

\section{Introduction}

The development of large-aperture telescopes and high-sensitivity
polarimetric instrumentation is providing the Astrophysical community
with new tools to explore the universe. Polarimetric observations are
opening new and exciting possibilities for the diagnostics of
light-matter interaction processes in many different scenarios (see,
e.g., the various reviews in \citeNP{WS02}). 

There are two major reasons why spectral lines produced in
stellar atmospheres may become polarized. First, we may have
scattering processes in which the photons emitted by the source are
deflected towards Earth. Depending on the details of the interaction,
one receives light that is (at least in part) linearly polarized in a
direction that is closely related to the scattering geometry (and some
times other factor, such as the magnetic field). This is obviously of
great interest for the detection and characterization of exoplanets
(e.g., \citeNP{K06}), protoplanetary and accretion
disks (e.g., \citeNP{DPB+05}; \citeNP{HGB+06}; \citeNP{PDK+06}),
binary systems (e.g., 
\citeNP{LdILdIL88}) etc.

Another source of polarized light is the Zeeman effect that splits the
atomic energy sublevels when magnetic fields are present in the
spectral line formation region. The strength of the magnetic field
required to produce a certain amount of polarization depends on
details of the radiative transfer (Zeeman sensitivity of the spectral
lines, opacity profile, source function gradient, etc). Angular
integration of a spherically-symmetric emitting source results in
circular polarization spectral profiles (linear polarization would
cancel out). Analysis of these spectral profiles provides unvaluable
clues to study, e.g. magnetic activity in stars (e.g.,
  \citeNP{WDB+05}; \citeNP{SGR+06}). 

For several decades now, the grid of radiative
equilibrium model atmospheres and spectra computed by Kurucz (for an
updated reference see \citeNP{CK03}) have been very useful in
the preparation and interpretation of stellar observations. We felt
that a polarization atlas of the signal emergent from those models,
for several magnetic field strengths and metallicities, would also be
useful for future observations of stellar magnetic fields. The entire
grid of polarized spectra computed in this work is publicly available
for download from the following URL:
\begin{verbatim}
http://download.hao.ucar.edu/pub/PSA/
\end{verbatim}

\section{The Atlas}

The synthetic spectra have been obtained using a compilation of
spectral line data from \citeN{KB95} between 400 and 1600~nm for
the model with $T_{\rm eff}$=5750~K and $\log(g)$=4.5 (hereafter, the
solar-like model), and between 400 and 1000~nm for the rest (if there
is enough interest, we may consider extending the wavelength
range). This is a useful range to consider, as it is accessible to
modern spectro-polarimetric instruments such as MUSICOS
(\citeNP{DCW+99}) or ESPaDOnS (\citeNP{MD03}).

Due to insufficient data on partition functions we restricted
ourselves to lines of up to the third ionization stage. Lines of
higher ionization stage have been excluded. The effective
Land\'e factor ($\bar g_{\rm eff}$) for lines that are not well described by
the $LS$ coupling or for which there is no level configuration
information has been set to zero, so no polarization signal will
appear in these lines\footnote{ Although we shall often refer to
effective Land\' e factors, our computation actually takes into
account the full Zeeman pattern (anomalous splitting) and does not
resort on the triplet approximation.  }. Following \citeN{CLdI94}, the
H$\alpha$ line is assumed to have an effective Land\'e factor of 1. A
comprehensive list of the lines excluded or computed with zero Land\'e
factor (35042 for the solar-like model and 27188 for the others) is
available for download in electronic format. The total number of lines
considered is 165501 in the extended range for the solar-like model
and 128794 for all others. Table~\ref{table:lines} lists the
distribution of spectral lines included and excluded in each
wavelength range.

The full polarized radiative transfer vector equation
for a given wavelength is usually written as:
\begin{equation}
\label{eq:RTeq}
{d{\bf I}(s) \over ds} = - {\bf K}(s) [{\bf I}(s) - {\bf S}(s)]\, ,
\end{equation}
where $s$ is the distance along the ray path, ${\bf I}$ is
the Stokes vector, ${\bf S}$ is the source function vector (which in
LTE is usually the Planck function times the transpose of the unitary
[1,0,0,0] vector), and finally ${\bf K}$ is the absorption matrix:
\begin{equation}
\label{eq:RT}
{\bf K}=\left ( \begin{array}{cccc}
   \eta_I & \eta_Q & \eta_U & \eta_V \\
   \eta_Q & \eta_I & \rho_V & -\eta_U \\
   \eta_U & -\rho_V & \eta_I & \rho_Q \\
   \eta_V & \rho_U & -\rho_Q & \eta_I \\
		\end{array} \right ) \, .
\end{equation}
For details on the meaning of the symbols employed in the equations
above, the reader is referred to \citeN{LdIL04}. Note that,
although we are only interested on the emergent intensity $I$ and
circular polarization $V$, we must compute all the four Stokes
parameters consistently due to the off-diagonal terms in ${\bf K}$. 

The syntheses were carried out using a modified version of the code
LILIA (\citeNP{SN01a}). This code was designed for the synthesis and
inversion of polarized atomic spectral lines formed in LTE (molecular bands
cannot be treated with this code). Solar metal abundances are those
from \citeN{AG89}. For all other metallicities, abundances are scaled by a
constant factor. Background opacities are obtained with the opacity
package used by \citeN{STB01}. The atmosphere is assumed to be
one-dimensional and plane-parallel. The formal solution of the
polarized transfer equation is based on the Hermitian method of
\citeN{BRRCC98}. However, a much simpler method may sometimes be used
(see the Weakly Polarizing Media approximation below). 

The plane-parallel approximation is justified by our choice of 
not considering the largest stellar models ($\log(g)$ between 3.5 and
5). The LTE approximation implicit in the expressions of the source
function and the absorption matrix is justified by the range of
effective temperatures, $T_{\rm eff}$ between 3500 and 10000~K.

For consistency, we started with the mass scale from the original models
and transformed it into an optical depth scale, using our opacities in
the calculation. The optical depths obtained are the reference height
scale for the radiative transfer. In order to avoid very hot (where
LTE may not be a good approximation for many photospheric lines) or
very large stars (where the spherical geometry may be needed), we
decided to restrict ourselves to a range of effective temperatures
(hereafter $T_{\rm eff}$) between 3500 and 10000~K and surface gravities
(hereafter $\log(g)$) between 10$^{3.5}$ and 10$^5$. The
mixing-length treatment of these models has been described by
\citeN{CGK97a} (see also \citeNP{CGK97b}).

We selected five different representative metallicities, with
abundance ratios (relative to solar) of 1, 10$^{-0.5}$, 10$^{-1}$,
10$^{-1.5}$, and 10$^{-2}$. For each abundance, $T_{\rm eff}$ and
$\log(g)$, we computed three different magnetic field strengths of
100~G, 1000~G and 5000~G. 

It is clear that the amount of all possible realizations of magnetic field
configurations on the surface of a star cannot be exhausted by any reasonable
grid of models; that is certainly not the aim of this work.
To find out the particular configuration that best fits some
observation, one would need to resort on least-squares inversion
methods (such as that of \citeNP{APGLL+00}). 
Our atlas has been calculated to give a qualitative account of actual
observations, as well as to explore possible observational strategies,
find the most promissing spectral lines for magnetic field
diagnostics, etc. Therefore, we have assumed the simplest possible
configuration, given by a monopolar magnetic field
pointing radially away from the center and of constant strength.

%For simplicity, the magnetic field is
%assumed to be homogeneous over the stellar surface and constant with
%height, {\bf with a radial geometry}. 
The spectra are calibrated in
flux using a 3-point Gaussian 
quadrature. We looked at several randomly-chosen models and verified
that increasing the quadrature grid to 4 or 5 points did not make any
significant difference.

\begin{table}
  \caption[]{Number of lines included and excluded in the syntheses}
  \label{table:lines}
  \begin{tabular}{rccc}
    \hline
    \noalign{\smallskip}
    Wavelength   &  Lines    &  Lines  &  Lines set \\
    range (nm)  & considered & excluded & to $\bar g_{\rm eff}$=0 \\ 
    \noalign{\smallskip}
    \hline
    \noalign{\smallskip}
    400.00-499.99  & 40343  & 835 & 8263 \\
    500.00-599.99  & 27849 & 421 & 6396 \\
    600.00-699.99  & 20371 & 192 & 4549 \\
    700.00-799.99  & 16141 & 215 & 2915 \\
    800.00-899.99  & 12793 & 104 & 2286 \\
    900.00-999.99  & 11297 & 92 &  2779 \\
    1000.00-1099.99  & 8601 & 43 &  2138 \\
    1100.00-1199.99  & 6821 & 67 &  1499 \\
    1200.00-1299.99  & 6016 & 54 &  1280 \\
    1300.00-1399.99  & 5357 & 26 &  1015 \\
    1400.00-1499.99  & 5153 & 17 &  948 \\
    1500.00-1599.99  & 4759 & 24 &  974 \\
    \noalign{\smallskip}  

    \hline
  \end{tabular}
\end{table}

%file0='excluded_'
%for i=0,11 do begin
% file=file0+string(i,format='(i3.3)')
% spawn,' grep -i ion '+file,lines
% excluded=n_elements(lines)-1
% spawn,'wc '+file,lines
% n=long((strsplit(lines(1),/extract))(0))-excluded
% print,file,':',n
%endfor
%end

The models used here have a depth-independent microturbulent velocity
of 2~km~s$^{-1}$ and a macroturbulence of 1~km~s$^{-1}$. These
are roughly representative solar values . In addition to turbulence,
other line-broadening mechanisms are taken into account after
\citeN{G76}:
\begin{itemize}
\item Radiative broadening. We set $\Gamma_r = 0.22233 / \lambda^2$
(with $\lambda$ in cm). This formula is only a rough approximation but
radiative broadening is usually very small anyway.
\item Van der Waals broadening. Note that the \citeN{G76} expressions
  are only accurate for neutral or singly ionized atoms. For other
  ions we neglect this broadening.
\item Stark broadening. We use the approximation $\Gamma_S\simeq$10.06666 +
  $\log_{10} (P_e)$-5/6 $\log_{10} (T)$, where $P_e$ and $T$ are the
  electron pressure and the local temperature (in c.g.s. units).
\end{itemize}

\begin{figure}
  \resizebox{\hsize}{!}{\includegraphics{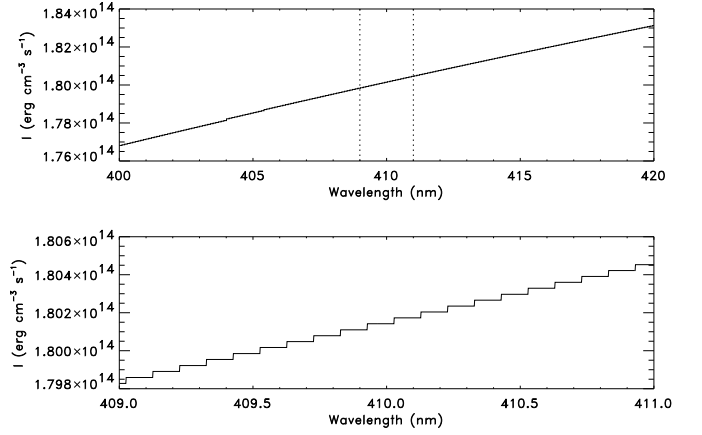}}
  \caption{Blue continuum spectrum for a solar-like model. The lower
  panel shows a close-up of the region marked by vertical dotted lines
  in the upper panel, rescaled to show the discretization of the
  continuum. Notice, however, in the ordinates scale that this effect
  is very small.}
  \label{fig:cont}
\end{figure}

\begin{figure}
  \resizebox{\hsize}{!}{\includegraphics{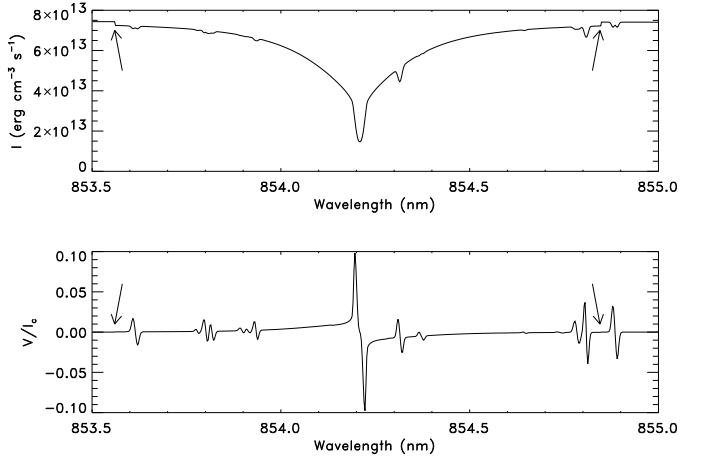}}
  \caption{Spectral profile of the \ion{Ca}{II} line in the solar-like
  model, showing some small glitches (marked by the arrows) in the
  line-to-continuum transition of Stokes~$I$ (upper panel). Stokes~$V$
  (lower panel) is much less sensitive to this effect.}
  \label{fig:8542}
\end{figure}

Even assuming LTE, the synthesis of the Stokes profiles of over 10${^5}$
spectral lines in a large grid of model atmospheres is a very
demanding computing task. To bring the computing time within
reasonable practical limits we implemented a number of optimizations
that do not harm significantly accuracty of the polarized spectra: 
\begin{itemize}
\item {\em Background opacities}. These are not recomputed at each
  wavelength. Instead, they are evaluated every 0.1~nm. This is
  noticeable, e.g. when one inspects closely the continuum intensity
  over a very short wavelength range. Figure~\ref{fig:cont} shows the
  continuum for the solar-like model. A very close zoom-in reveals the
  discretization in the continuum caused by this
  approximation. Notice, however, the ordinates scale which shows that
  the continuum is virtually flat and thus the error introduced (the
  ``step'' of the staircase) is
  very small, at the 10$^{-4}$ level.
\item {\em Weak lines}. Depending on the particular physical
  conditions on a given 
  atmosphere, many spectral lines turn out to be negligibly
  weak. Removing these lines from the computation saves a large amount
  of processing power. We implemented a criterion, according to which
  a line is ignored if its line-core opacity is at most 10\% larger
  than the continuum opacity at all depth-points in the
  atmosphere. The line-core opacity is calculated as:
\begin{equation}
	  \eta_0=4.9947\times 10^{-21} { 10^{\log(g_l f)} (n_l - n_u)
          \over \rho \Delta\lambda_D^2} \, ,
\end{equation}
where $g_l$ is the statistical weight of the lower level, $f$ is the
oscillator strength, $n_l$ and $n_u$ are the atomic lower and upper
level number densities, $\rho$ is the gas density and
$\Delta\lambda_D$ is the Doppler width (see Eq~[\ref{eq:doppler}]
below). If all these parameters are given in c.g.s units, the
resulting $\eta_0$ is in units of cm$^2$ g$^{-1}$.
\item {\em Line profiles}. Computing Voigt functions (\citeNP{H82})
  for a large number of 
  lines, wavelength points and many Zeeman components is the dominant
  fraction of the spectral synthesis time, even when using an
  optimized vector routine. Extending the Voigt profiles too far into
  the wings results in a lot of unnecessary processing. On the other
  hand, falling short produces inaccuracies near the continuum. A good
  estimate of the width of the line is therefore very important in
  optimizing the efficiency of the code. We have implemented an
  adaptive scheme that starts with a standard width of 0.01~nm and
  extends the interval until the opacity at the boundaries is smaller
  than 1\% of the continuum opacity. Figure~\ref{fig:8542} shows the
  \ion{Ca}{II} line at 854.2~nm in the solar-like model. This is
  perhaps the line that shows the most conspicuous glitch (marked by
  arrows in the figure) in the transition from line to continuum. The
  glitch is noticeable only marginally in Stokes~$I$ and is negligible
  in Stokes~$V$.
\item {\em Wavelength sampling}. Since the Doppler width of the lines
  increases with 
  wavelength (and the Zeeman splitting increases with the square of
  the wavelength), one needs more resolution in the blue than in the
  red side of the spectrum. We divided the spectra in 100~nm intervals
  and used a sampling typical of solar observations in the visible
  ($\simeq$20~m\AA \, at 650~nm). This sampling is then scaled
  linearly with wavelength. It may be seen that even for cool
  atmospheres and relatively heavy ions our sampling is still smaller
  than the Doppler width of the lines, given by:
\begin{equation}
\label{eq:doppler}
   \Delta\lambda_D = \lambda \left ( \sqrt{ {2 k T \over m_A } } + 
         {v_{mic} \over c} \right ) \, ,
\end{equation}
where $k$ is the Boltzmann constant, $T$ is the temperature in the
atmosphere, $m_A$ is the atomic mass, $v_{mic}$ is the microturbulence
and $c$ is the speed of light.
\item {\em Weakly Polarizing Media}. In many cases, the radiation
  field is weakly linearly polarized, while $\eta_{Q, U}/\eta_I \ll
  1$, and magneto-optical terms are negligible ($\rho_{Q, U, V}/\eta_I
  \ll 1$). In that case, the coupled system of Eq~(\ref{eq:RTeq})
  simply reduces to two decoupled differential equations for $I\pm V$
  that are easily integrated. If circular polarization is also weak,
  we can neglect the $-\eta_V V$ from the equation for the intensity
  (this is the Weakly Polarizing Media approximation of \citeNP{SATB99})
  and we get an independent equation for the intensity $I$, and an
  easily integrable equation for $V$. If 
  the criterion above is satisfied for all depth-points in the
  atmosphere, the code automatically switches to this method for the formal
  solution. 
\end{itemize}

\section{Discussion}

\begin{figure}
  \resizebox{\hsize}{!}{\includegraphics{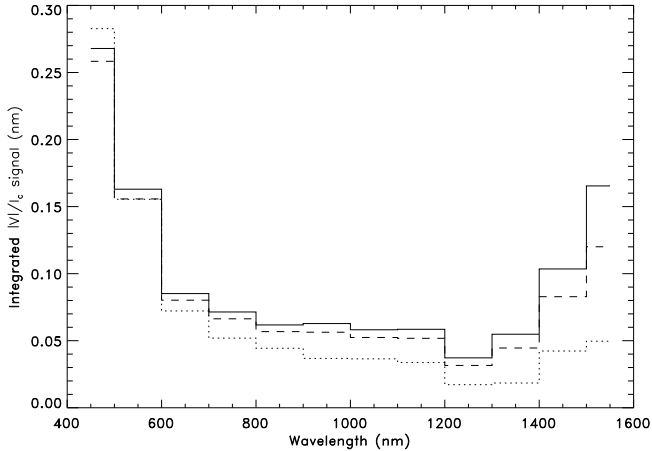}}
  \caption{Integrated absolute value of Stokes~$V$ (in units of
  the continuum intensity) over 100~nm bins of the solar-like
  model. Solid: 100~G. Dashed: 1000~G. Dotted: 5000~G. The scale of
  the dashed (dotted) curve has been reduced by a factor of 10 (30)
  in order for all three curves to be plotted together.}
  \label{fig:totv_p00}
\end{figure}

Because the Zeeman splitting increases with wavelength, one might
expect infrared regions to be more promising targets for magnetic
field observations. However, this is not necessarily the
case. Figure~\ref{fig:totv_p00} shows the total amount of polarization
(in absolute value) measured in each 100~nm interval of the solar-like
model spectrum, i.e., $\int_{\lambda_0}^{\lambda_1} {\rm
|}V(\lambda){\rm |}/I_c(\lambda) d\lambda$. By far, the blue side of
the spectrum carries a much higher density of polarization signal than
the red or infrared. In fact, for the strongest fields, the region
from 400 to 500~nm exhibits nearly as much signal as the rest of the
spectrum combined (from 500 to 1600~nm). The reason for this is
two-fold. First, the blue region is heavily crowded with a large
number of closely packed spectral lines. This is evident from
Table~\ref{table:lines}, where we can see that the number of lines per
wavelength interval decreases monotonically towards longer
wavelengths. Second, blue lines are typically much stronger than red
or infrared lines. The degree of polarization produced by a
Zeeman-split line depends not only on the magnetic field and the
component splitting, but also on the line strength.

We can see in Figure~\ref{fig:totv_p00} the balance among all the
different effects for this model. The amount of Stokes~$V$ signal is
largest at 400~nm and decreases monotonically across the visible
spectrum. A minimum is reached between 1200 and 1300~nm, after which
the three curves start to increase again all the way up to
1600~nm. For the stronger fields, the signal drop with wavelength is
considerably more pronounced and the increase after 1300~nm is much
weaker. This is because the amount of polarization produced by a line
increases with the field strength only in the so-called {\it
weak-field regime} (see, e.g., \citeNP{LdI92}). This is the regime in
which the Zeeman splitting is smaller than the Doppler width of the
line. If all other parameters are constant, the ratio of Zeeman
splitting to Doppler width increases linearly with wavelength. As we
increase the field strength, red and infrared lines are more likely to
depart from the weak-field regime before blue lines do. In the
saturation regime, the polarization amplitude does not increase with
the magnetic field any more. In summary, observing longer wavelengths
is only advantageous (in terms of polarization amplitude) for
relatively weak fields ($\sim$100~G).

If the magnetic field is not homogeneous and covers only a fraction of
the stellar surface, then the Stokes~$V$ spectrum will have its
overall amplitude decreased according to the {\it filling factor}
(fractional surface area occupied by the magnetic field). Generally,
it is not straightforward to separate the field strength from its
filling factor. Figure~\ref{fig:totv_p00} gives us a clue on how this
might be accomplished, since the shapes of the curves for different
intrinsic field strengths are clearly different. If one were to take
the ratio of polarization signal in the 800~nm to the 400~nm band,
this ratio would be larger for the weaker fields and decrease as the
field becomes stronger. This is a similar concept to the line-ratio
technique used by \citeN{S73} to propose that the solar magnetic
network is actually composed of strong kG fields occupying small
filling factors.

There are many other lines that could potentially be used to
disentangle intrinsic field strength from filling
factor. Table~\ref{table:identical} lists a number of spectral lines
that have the same line opacity (same lower level excitation potential
and $\log(gf)$). The \ion{Fe}{I} lines at 412.180 and 899.956~nm are
particularly promising (see \citeNP{SNBAR+06}), although the disparity
in wavelength may introduce complications (most notably, the
background opacities are sensibly different). There is a number of
\ion{Fe}{I}, \ion{Ca}{I}, \ion{Ti}{I} and \ion{Si}{I} that also have
the same Land\' e factor, and are therefore virtually identical except
for their wavelength. These lines may be useful, for example to
determine accurate atmospheric temperatures through the calculation of
their equivalent width ratio.

\begin{figure}
  \resizebox{\hsize}{!}{\includegraphics{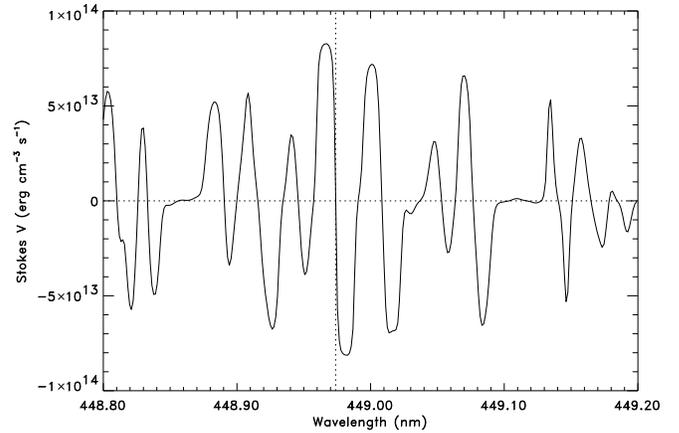}}
  \caption{Stokes~$V$ signal in the region around 449~nm. This is
  where the strongest polarization is found for the 5000~G field. The
  vertical dotted line marks the central wavelength of a strong
  \ion{Fe}{I} transition at 448.9739~nm. The half-amplitude of this profile
  is approximately 44\% of the continuum intensity.}
  \label{fig:strongblue}
\end{figure}

Table~\ref{table:strongest} lists the wavelengths with strongest
polarization signal in the solar-like model. We can see that for weak
fields the strongest signals are in the infrared. As the field
strength increases, most of the infrared lines saturate and the
strongest signals are observed in shorter wavelengths. Note that it is
not straightforward in general to associate specific spectral lines to
those wavelengths because some times (especially in the blue), the
polarization signal arises from a blend of several lines. For example,
the strongest signal in the solar spectrum produced by weak fields
($\sim$100~G) comes from a \ion{Fe}{I} line with central wavelength at
1159.358~nm with an effective Land\'e factor of 2.5. For very strong
fields, on the other hand, the strongest signal comes from a very
crowded region around 448.97~nm. Figure~\ref{fig:strongblue} shows the
Stokes~$V$ profile in this region. The most conspicuous feature is
produced mostly by an \ion{Fe}{I} line at 448.9739~nm (vertical dotted
line), but there are other lines with very close wavelengths that
contribute to it.

\begin{figure*}
  \resizebox{\hsize}{!}{\includegraphics{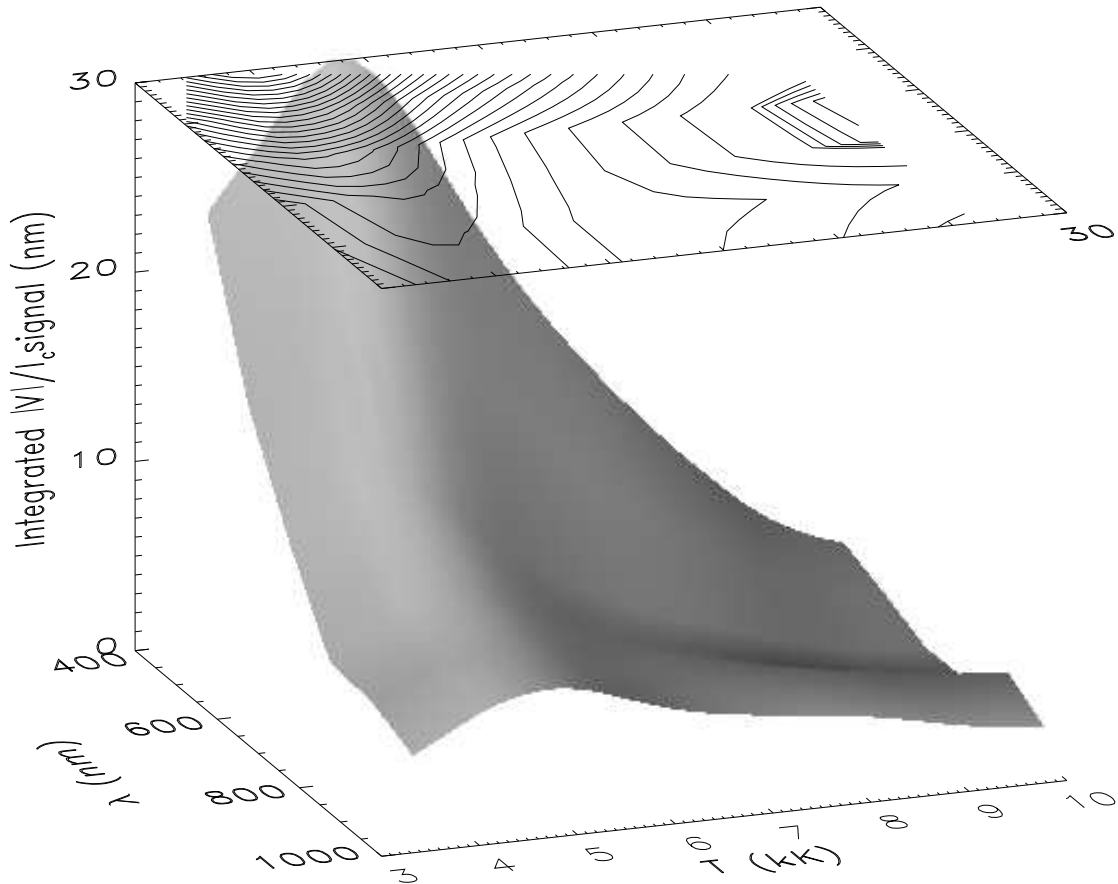}}
  \caption{Integrated absolute value of Stokes~$V$ (in units of
  the continuum intensity) over 100~nm bins for varying spectral
  types and a 1~kG field.}
  \label{fig:surfa}
\end{figure*}

The variation with spectral type is shown in Fig~\ref{fig:surfa} for
$\log(g)$=4.5. The later spectral types (left of the figure) have a
very prominent blue side of the spectrum, with a steep decrease
towards the red. Earlier types exhibit a rather flat spectral
dependence, with a minimum around 700~nm. Notice also that there is an
overall decrease of the polarization signal for hotter stars. This is
due to a combination of three different physical effects. First, the
thermal broadening of spectral lines increases with $T_{eff}$,
smearing out the polarization signal. Second, the increase in
excitation and ionization state of the atoms in the plasma reduces the
number and strength of spectral lines (especially, the abundant
low-excitation \ion{Fe}{1} lines). This effect is particularly
important in the blue. Finally, if we compute the source function
($S$) and its gradient ($dS/d\tau$) and plot the ratio $(dS/d\tau)/S$
in the photosphere as a function of $T_{eff}$, we find that this ratio
has a maximum around 5000~K and then decreases monotonically for
hotter stars.

\begin{table}
  \caption[]{Compilation of spectral line pairs with identical line opacities}
  \label{table:identical}
  \begin{tabular}{lcccc}
    \hline
    \noalign{\smallskip}
         Ion & Wavelength   &  Excitation  &  $\log(gf)$ & $\bar g_{\rm eff}$ \\
             &   (nm)       & potential (eV) &             &        \\
    \noalign{\smallskip}
    \hline
    \noalign{\smallskip}
\ion{Fe}{I} &   400.560 &    4.186 &   -2.131 &   2.0000 \\
\ion{Fe}{I} &   453.894 &    4.186 &   -2.131 &   0.8333 \\
\hline
\ion{Fe}{I} &   402.473 &    3.241 &   -0.710 &   0.6250 \\
\ion{Fe}{I} &   408.530 &    3.241 &   -0.710 &   1.7083 \\
\hline
\ion{Ti}{I} &   406.509 &    1.053 &   -0.760 &   1.5000 \\
\ion{Ti}{I} &   469.876 &    1.053 &   -0.760 &   1.0000 \\
\hline
\ion{V}{I} &   410.516 &    0.267 &   -0.130 &   1.5000 \\
\ion{V}{I} &   440.851 &    0.267 &   -0.130 &   1.4667 \\
\hline
\ion{Fe}{I} &   412.180 &    2.832 &   -1.300 &   0.6667 \\
\ion{Fe}{I} &   899.956 &    2.832 &   -1.300 &   1.5000 \\
\hline
\ion{Na}{I} &   421.606 &    2.104 &   -2.350 &   1.1000 \\
\ion{Na}{I} &   454.518 &    2.104 &   -2.350 &   1.1667 \\
\hline
\ion{Fe}{I} &   428.582 &    3.640 &   -1.780 &   1.1667 \\
\ion{Fe}{I} &   496.870 &    3.640 &   -1.780 &   0.8333 \\
\hline
\ion{Ca}{I} &   435.508 &    2.709 &   -0.430 &   1.0000 \\
\ion{Ca}{I} &   452.693 &    2.709 &   -0.430 &   1.0000 \\
\hline
\ion{Fe}{I} &   443.834 &    3.686 &   -1.630 &   2.5000 \\
\ion{Fe}{I} &   466.806 &    3.686 &   -1.630 &   1.7500 \\
\hline
\ion{Mn}{I} &   445.301 &    2.941 &   -0.490 &   1.5000 \\
\ion{Mn}{I} &   473.909 &    2.941 &   -0.490 &   0.8000 \\
\hline
\ion{Fe}{I} &   456.531 &    3.274 &   -2.510 &   2.0000 \\
\ion{Fe}{I} &   479.125 &    3.274 &   -2.510 &   0.5000 \\
\hline
\ion{Fe}{I} &   457.982 &    3.071 &   -2.830 &   1.7500 \\
\ion{Fe}{I} &   718.915 &    3.071 &   -2.830 &   1.0000 \\
\hline
\ion{Fe}{I} &   461.929 &    3.603 &   -1.120 &   1.8333 \\
\ion{Fe}{I} &   463.801 &    3.603 &   -1.120 &   1.7917 \\
\hline
\ion{Fe}{I} &   463.585 &    2.845 &   -2.420 &   2.2500 \\
\ion{Fe}{I} &   635.503 &    2.845 &   -2.420 &   1.0000 \\
\hline
\ion{Ti}{I} &   469.133 &    1.067 &   -0.840 &   1.1667 \\
\ion{Ti}{I} &   586.645 &    1.067 &   -0.840 &   1.1667 \\
\hline
\ion{Fe}{I} &   480.065 &    4.143 &   -1.260 &   1.0000 \\
\ion{Fe}{I} &   518.792 &    4.143 &   -1.260 &   1.0000 \\
\hline
\ion{Si}{I} &   540.466 &    5.614 &   -0.530 &   2.0000 \\
\ion{Si}{I} &   548.898 &    5.614 &   -0.530 &   2.0000 \\
\hline
\ion{Si}{I} &   540.466 &    5.614 &   -0.530 &   2.0000 \\
\ion{Si}{I} &   623.732 &    5.614 &   -0.530 &   2.0000 \\
\hline
\ion{Si}{I} &   540.985 &    5.616 &   -0.690 &   2.0000 \\
\ion{Si}{I} &   624.447 &    5.616 &   -0.690 &   2.0000 \\
\hline
\ion{Si}{I} &   548.898 &    5.614 &   -0.530 &   2.0000 \\
\ion{Si}{I} &   623.732 &    5.614 &   -0.530 &   2.0000 \\
\hline
\ion{Fe}{I} &   560.294 &    3.430 &   -0.851 &   0.7500 \\
\ion{Fe}{I} &   565.853 &    3.430 &   -0.851 &   2.2500 \\
\hline
\ion{Si}{I} &   565.492 &    5.614 &   -0.930 &   2.0000 \\
\ion{Si}{I} &   612.502 &    5.614 &   -0.930 &   2.0000 \\
    \noalign{\smallskip}  
    \hline
  \end{tabular}
\end{table}

\begin{table}
  \caption[]{Wavelengths (in nm) exhibiting strongest polarization in
  the solar-like model. The numbers in parenthesis give the
  degree of polarization (relative to the local continuum).}
  \label{table:strongest}
  \begin{tabular}{ccc}
    \hline
    \noalign{\smallskip}
         100~G    &  1000~G  &  5000~G \\
    \noalign{\smallskip}
    \hline
    \noalign{\smallskip}
    1159.34 (7.18\%) &  525.02 (35.38\%) &    448.97 (44.36\%)\\
    846.83 (6.56\%) &   423.27 (34.63\%) &    420.66 (44.13\%)\\
    1449.77 (6.27\%) &   425.83 (34.21\%) &   411.17 (43.63\%)\\
    525.02 (5.94\%) &   524.75 (33.64\%) &    431.28 (43.37\%)\\
    1564.84 (5.78\%) &   408.08 (33.62\%) &   400.16 (43.24\%)\\
    1386.41 (5.74\%) &   513.14 (33.36\%) &   417.49 (43.11\%)\\
    1168.98 (5.71\%) &   461.33 (33.33\%) &   408.29 (43.02\%)\\
    1142.22 (5.45\%) &   454.59 (31.96\%) &   412.12 (42.67\%)\\
    1456.57 (5.45\%) &   550.67 (31.95\%) &   429.14 (42.65\%)\\
    513.14 (5.38\%) &   522.55 (31.89\%) &    457.10 (42.55\%)\\
    524.75 (5.37\%) &   448.56 (31.66\%) &    437.44 (42.55\%)\\
    525.01 (5.26\%) &   429.14 (31.10\%) &    438.92 (42.53\%)\\
    1399.74 (5.25\%) &   465.44 (30.81\%) &   405.55 (42.43\%)\\
    \noalign{\smallskip}  

    \hline
  \end{tabular}
\end{table}

\section{Conclusions}

Stellar polarimetry is becoming increasingly important in Astrophysics
and it is now possible to obtain polarized spectra over a broad range
of wavelengths (examples are the Narval and Musicos instruments at Pic
du Midi, \citeNP{A03}; or ESPaDOnS at the Canada-France-Hawaii
Telescope, \citeNP{PDE03}), or over a narrower range with very high
spatial resolution and large aperture (FORS at the VLT; e.g.,
\citeNP{HSY04}; \citeNP{HKB+04}).

We have used a Stokes synthesis code and a large array of LTE models
to produce an atlas of the polarization signal expected from a wide
variety of magnetic stars. We hope that this tool will be useful for
the planning of future polarimetric observing runs and perhaps for the
interpretation of the data, as well. 

For example, it may be possible to (coarsely) generalize the popular
least-squares deconvolution technique (\citeNP{DSC+97}) to
intermediate and strong magnetic regimes. In order to do this, one
would take the atlas for the specific star under consideration and add
all the spectral lines together, as is usually done when applying
least-squares deconvolution, for different magnetic field
strengths. By comparing the observed profile to those obtained from
the atlas, we may be able to estimate the stellar magnetic field.

In principle, it is conceivable to use the synthesis module developed
here for the application of least-squares inversions (e.g.,
\citeNP{SN01a}; \citeNP{dTI03}) to stellar spectro-polarimetric
observations. It would be a siginifcant computational undertaking,
however, since this kind of inversions typically require of the order
of $\sim$100 syntheses (or, if one wishes to use numerical response
functions, this figure would have to be multiplied by the number of
free parameters). An interesting alternative may be the use of
inversions based on Principal Component Analysis combined with a
look-up algorithm (\citeNP{RLAT+00}; \citeNP{SNLAL01};
\citeNP{SRRV+06}). To this aim, 
it would be desirable to develop a more refined grid with a better
sampling of the magnetic field and, possibly, a variety of geometries.

A final caveat in using our atlas is that it has been derived from a
very simple magnetic toplogy. Active stars are likely to exhibit more
complex magnetic structuring. The integration of different field
orientations over the stellar disk would result in an overall decrease
of the circular polarization signal with respect to the synthetic
profiles presented here. However, the relative scaling between
different lines and spectral regions should be approximately correct.

\begin{acknowledgements}
This work has been partially funded by the Spanish
Ministerio de Educaci\'on y Ciencia through project
AYA2004-05792
\end{acknowledgements}

\bibliographystyle{../bib/apj}
\bibliography{../bib/aamnem99,../bib/articulos}

\end{document}